\documentclass[cameraready]{Interspeech}
\usepackage{hyperref} 
\usepackage{comment}  
\usepackage{cite}
\usepackage{amsmath,amssymb,amsfonts}
\usepackage{algorithmic}
\usepackage{float}
\usepackage{xcolor,colortbl} 
\usepackage{amsmath,amsfonts} %
\usepackage{graphicx}%
\usepackage{textcomp}%
\usepackage{xcolor}%
\usepackage{url}
\usepackage{array}%
\usepackage{pifont}%

\usepackage{amsthm}%
\usepackage{bm}

\usepackage[linesnumbered,ruled,vlined]{algorithm2e}%
\usepackage{setspace}%

\usepackage{multirow}%
\usepackage{siunitx}
\usepackage{footnote}%
\usepackage[normalem]{ulem}
\usepackage{soul}

\title{CS-ETS: \underline{C}haos-Inspired \underline{S}amba-Based \underline{E}MG-\underline{T}o-\underline{S}peech Synthesis with Nonlinear Chaotic Losses}

\author{Sajid Fardin}{Dipto}
\author{Tarikul Islam}{Tamiti}
\author{David}{Vergano}
\author{Luke}{Baja-Ricketts}
\author{Anomadarshi}{Barua}

\address{\mbox{}}

\email{}


\usepackage{comment}
\usepackage{booktabs}
\begin{document}

\maketitle

\begin{abstract}
\vspace{-0.5em}

    We propose a chaos-inspired new architecture for EMG-to-Speech (ETS) synthesis called CS-ETS, which combines a Samba-based encoder with two novel chaos-inspired loss functions -- Lyapunov Exponent Regularization (LER) and Multi-Scale Detrended Fluctuation Analysis (MSDFA). LER is designed based on Lyapunov exponents to capture nonlinear fluctuations and sensitivity to initial conditions. MSDFA exploits detrended fluctuation analysis to quantify fractal-like, long-range temporal chaotic correlation. CS-ETS surpasses prior work with a 40.79\% lower parameter count (32M vs 54.1M) and introduces a new Post-Vocoder Alignment approach that improves LSD by 2.1x, STOI by 4.7x, and SI-SDR by 1.25x. CS-ETS reduces computation by 13.33\% while maintaining improved performance. To the best of our knowledge, for the first time, we show how ETS can be supervised by the subtle non-linear chaotic physics with Samba attention to achieve a significantly smaller model with superior performance.
    
\end{abstract}

\vspace{-0.8em}
\section{Introduction}
\vspace{-0.4em}
Silent speech interfaces (SSIs) allow speech synthesis without vocalization by interpreting articulatory bio-signals \cite{satterlee2025}, supporting individuals with speech impairment and enabling communication in acoustically challenging environments \cite{dong2023}. Electromyography (EMG) is an effective modality for recording articulatory muscle activity for SSI speech production \cite{satterlee2025, kurotaki2025}.

\textbf{EMG as a chaotic dynamical system:} 
Articulatory EMG captures coordinated activation of facial and laryngeal muscles during speech production \cite{gaddy-klein-2020-digital, gaddy-klein-2021-improved} that can be modeled as \textit{deterministic chaos} due to \textbf{(i)} non-stationary motor unit firing patterns varying with force, fatigue, and phonetic context \cite{enoka2001, farina2015}, and \textbf{(ii)} temporal summation of asynchronous motor unit action potentials creating amplitude modulation and quasi-periodic bursts \cite{keenan2005}. Two critical chaotic features, such as \textbf{(i)} recurrent activation patterns across phonetic contexts reflecting articulatory gestures \cite{saltzman1989}, and \textbf{(ii)} rapid state transitions during coarticulation \cite{tuller1984}, can be estimated through \textit{Detrended Fluctuation Analysis} \cite{peng1994mosaic} and \textit{Lyapunov Exponents} \cite{oseledec1968multiplicative}. These multi-scale temporal dependencies encode subtle articulatory transitions, co-activation patterns, and speaker-specific muscle coordination essential for accurate phonetic decoding in EMG-to-Speech (ETS)  conversion \cite{gaddy-klein-2020-digital, gaddy-klein-2021-improved}.

\textbf{What current ETS systems miss:} Prior works \cite{gaddy-klein-2020-digital, gaddy-klein-2021-improved, scheck2023} face two critical limitations, despite progress in ETS synthesis. \textit{\ul{First}}, they only use reconstruction losses (L1, MSE) that minimize frame-level differences, but ignore the \textit{nonlinear chaotic dynamics} inherent to ETS synthesis. EMG signals exhibit deterministic chaos with sensitivity to initial conditions, yet training objectives in previous methods completely ignore these chaotic properties. \textit{\ul{Second}}, evaluation is limited to word error rate (WER) because post-vocoder temporal misalignment prevents frame-level comparison that makes standard acoustic metrics (LSD, STOI, PESQ) unattainable to compute. Perceptual quality thus remains uncharacterized beyond transcription accuracy.


\textit{To the best of our knowledge, our proposed Chaos-inspired Samba-based ETS (CS-ETS) is the first model to explicitly incorporate nonlinear chaos theory to state space models (SSMs) with a significant reduction in the parameter count.} We introduce two chaos-inspired losses: \textbf{LER}, which penalizes mismatches in local divergence rates by aligning Lyapunov Exponents across different resolutions; and \textbf{MSDFA}, which exploits detrended fluctuation analysis to quantify fractal-like, long-range temporal chaotic correlation.  Our contributions include: \textbf{\ul{(1)}} We introduce \textit{``modified Samba-based SSMs''} with sliding window attention, resulting in a 50.1\% reduction in encoder parameters compared to \cite{gaddy-klein-2021-improved}. \textbf{\ul{(2)}} We introduce post-vocoder alignment, enabling comprehensive acoustic evaluation (LSD, STOI, PESQ, SISDR, NISQA-MOS, see Section \ref{subsec:metrics} for full form) beyond WER for the first time in ETS.  CS-ETS delivers the 2.1x best spectral reconstruction (LSD = 1.07 vs 2.25), 4.7x improvement in intelligibility (STOI = 0.61 vs 0.13), and 1.25x improvement in noise reduction (SI-SDR = -33.41 vs -41.96) compared to baseline  \cite{gaddy-klein-2021-improved}. \textbf{\ul{(3)}} We introduce chaos-inspired LER and MSDFA losses with \textit{``modified Samba-based SSMs''}, enabling a 40.79\% reduction in total parameter count with significantly improved performance on the Gaddy dataset \cite{gaddy-klein-2020-digital}, supporting that chaotic supervision yields better performance with reduced model size.

{\color{red}

}

\begin{figure*}[t] 
    \centering
    \includegraphics[width=1.0\textwidth]{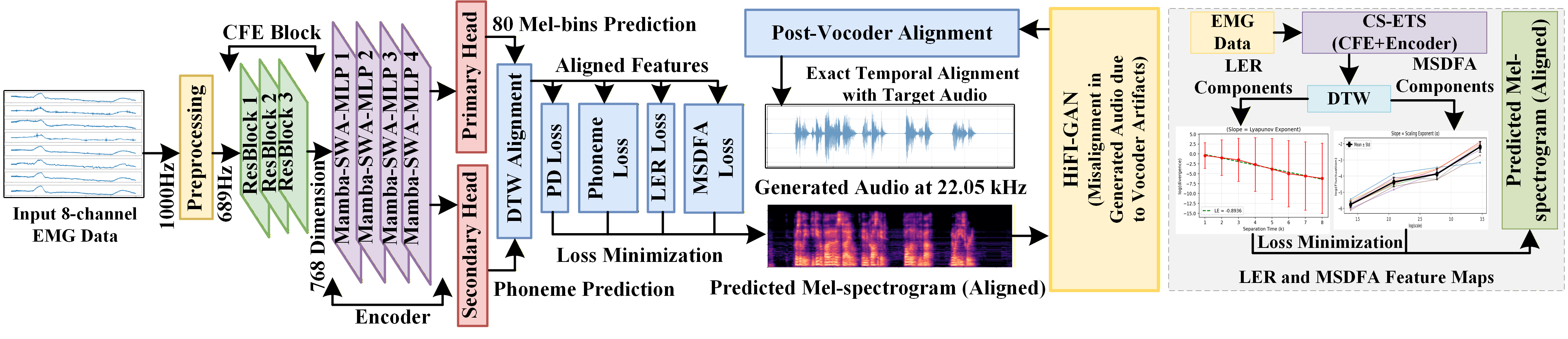} 
    \vspace{-02.4em}
    \caption{Detailed Architecture of CS-ETS Model with Chaos-Inspired Nonlinear Dynamics Losses Combined with Modified Samba.}
    \label{fig:arch}
    \vspace{-2.0em}
\end{figure*}

\vspace{-0.8em}
\section{ Chaos-Inspired Architecture Design}
\label{sec:architcture_design}
\vspace{-0.4em}

\textbf{Why we design chaos-inspired models:} Speech production is fundamentally a \emph{non-linear dynamical process characterized by deterministic chaos} \cite{little2007exploiting, tamiti2025nldsibwenonlineardynamical, tamiti2025nonlinearframeworkspeechbandwidth}. Therefore, ETS synthesis is inherently chaotic because it attempts to reconstruct a fundamentally non-linear, feedback-driven speech production system from partial muscular observations. 
\textit{Small variations in EMG signals can therefore produce disproportionately large acoustic changes, a hallmark of deterministic chaos. Non-chaotic models smooth out these effects, eliminating jitter, turbulence, and aperiodic bursts and yielding unnaturally clean speech.} \textit{Accurate EMG-to-speech synthesis thus requires non-linear chaos-inspired models that can represent both structured harmonic content and chaotic, irregular dynamics.}

CS-ETS has the following contributions in its architecture:
\vspace{-0.9em}

\begin{itemize}

\item CS-ETS is the \textit{first work}, which is based on a language model with efficient unlimited context called Samba \cite{samba2024}. We modify the Mamba-SWA-MLP framework along with a Convolutional Feature Extraction (CFE) block at the top layer to support chaotic feature extraction in the loss functions.

\item CS-ETS is the \textit{first work}, which includes \textit{chaos-inspired losses} to apprehend diverse chaotic and non-linear temporal cues \cite{chaos2025} in addition to the auxiliary phoneme loss \cite{gaddy-klein-2021-improved}. 

\end{itemize}

The architecture of CS-ETS is depicted in Fig. \ref{fig:arch}. It takes 8-channel EMG data as input, preprocesses it, and generates melspectrogram (mel) bins that are converted to audio at the last stage by a fine-tuned HiFi-GAN vocoder \cite{hifigan2020}. 

\vspace{-0.7em}
\subsection{Convolutional Feature Extraction (CFE) Block}
\label{subsec:CFEblock}
\vspace{-0.4em}


The CFE employs three residual blocks adapted from \cite{gaddy-klein-2021-improved}, where each block contains two one-dimensional convolutional (Conv-1D) layers (kernel=3) with ReLU activations, plus a residual connection. All layers maintain channel dimension, $C_D$=768. \textit{This configuration helps to capture dynamic facial patterns, achieving better performance with less parameters (see Table \ref{tab:ablation3} for channel ablation).}

\vspace{-0.7em}
\subsection{Mamba-SWA-MLP for Efficient Unlimited Context}
\label{subsec:Mamba-SWA-MLP}
\vspace{-0.4em}
Our encoder comprises Mamba \cite{mamba2024}, Sliding Window Attention (SWA) \cite{swa2020}, and Multi-Layer Perceptron (MLP), forming a \textit{modified} Samba architecture \cite{samba2024}. We introduce the following modifications to improve efficiency: (i) removed the first MLP sublayer \cite{samba2024} (reducing MLP-Mamba-SWA-MLP to Mamba-SWA-MLP) to make the encoder lightweight, (ii) replaced rotary positional embeddings (RoPE) in SWA of Samba \cite{samba2024} with relative positional embeddings from \cite{gaddy-klein-2021-improved} to capture temporal invariance, and (iii) used four Mamba-SWA-MLP layers for optimal performance \textit{(see Table \ref{tab:ablation4} for layer count ablation)}.

Following the CFE frontend, the extracted features are sent to the encoder which captures long-range temporal dependencies through SSMs \cite{mamba2024, samba2024} and computes scaled dot-product attention scores with a SWA mask \cite{swa2020}. After all Mamba-SWA-MLP layers, the model outputs through two heads: a primary linear projection (dimension 80) for melspectrogram generation, and an auxiliary head for phoneme-level supervision \cite{gaddy-klein-2021-improved}.

\vspace{-0.7em}
\subsection{Chaos-Inspired Loss Functions and  Alignment}
\vspace{-0.4em}
For the first time, we propose Lyapunov Exponent Regularization (LER) and Multi-scale Detrended Fluctuation Analysis (MSDFA) losses derived from chaos theory, explained below, in addition to the Dynamic Time Warping (DTW), Pairwise Distance (PD) loss, and phoneme loss \cite{gaddy-klein-2021-improved}.

\vspace{-0.7em}
\subsubsection{\textbf{Lyapunov Exponent Regularization (LER)}}
\vspace{-0.4em}

We introduce Lyapunov Exponent (LE) \cite{oseledec1968multiplicative, wolf1985determining} based LER loss to capture rapid, nonlinear fluctuations and sensitivity to initial conditions in speech that previous works overlook. The LER is a measure of nonlinear dynamics used to quantify the rate of separation of infinitesimally close trajectories that drives the model to reproduce deterministic chaotic behavior, yielding more lifelike and temporal stability in predicted audio features.

The largest LE is computed on non-overlapping windows with time-delay embedding dimension, m=10, and time delay, $\tau$=1 (\ul{see Alg.}\ref{alg:MRLD}). The extracted features are first embedded into a higher-dimensional phase space using time-delay embedding, $E_V[i] = [x[i, :], x[i + \tau, :], \dots, x[i + (m - 1)\tau, :]] \in \mathbb{R}^d$ where $x \in \mathbb{R}^{T \times C}$ is the input feature sequence with $T$ time frames and $C$ feature channels, and $d = m \cdot C$ is the embedding dimension (\textbf{Rows \textcircled{\scriptsize 1}--\textcircled{\scriptsize 4}}). Then, we compute the Positive Definite (PD) matrix, $D[i, j]$=$\big\Vert E_V[i] - E_V[j]\big\Vert_2$ for $i, j \in \{0, 1, \ldots, M-1\}$. Here, the maximum number of embedding vectors to use is $M$=$T-(m-1)\tau$ (\textbf{Rows \textcircled{\scriptsize 5}--\textcircled{\scriptsize 7}}). Then, we find the nearest neighbor for each point, $j^*(i) = \arg \min_{j \in \{1, \dots, M\}} D[i, j]$ while excluding temporally correlated neighbors using Theiler window \cite{Theiler1986} and track the distance evolution for each time offset $k \in \{1, 2, \dots, k_{max}\}$, $d_k[i]$=$\big\Vert E_V[i + k] - E_V[j^*(i) + k]\big\Vert_2$ (\textbf{Rows \textcircled{\scriptsize 9}--\textcircled{\scriptsize 12}}). Then, we compute the logarithmic divergence of nearest neighbor pairs over time for each $k$, $S(k) = \frac{1}{N_k} \sum_{i=1}^{M} \log (d_k[i] + \epsilon_1)$, where $\epsilon_1$=$10^{-8}$ prevents logarithm of zero (\textbf{Row \textcircled{\scriptsize 14}}). The largest LE is defined as, \small $\lambda = (\sum_{i=1}^{n_k} \tilde{k}_i \tilde{S}_i) \big/ (\sum_{i=1}^{n_k} \tilde{k}_i^2 + \epsilon_2)$ where $\epsilon_2$=$10^{-8}$ prevents division by zero. $\tilde{k}$ and $\tilde{S}$ represent the centered values of time offsets and logarithmic divergences, respectively. Then, we define LER loss as $\mathcal{L}_{\text{LER}}(\hat{Y}_S[j], Y_V[i])$=$\big\vert \lambda^{(\hat{Y}_S[j])} - \lambda^{(Y_V[i])}\big\vert$, where $\hat{Y}_S[j]$ and $Y_V[i]$ denote temporally aligned predicted and target audio features, respectively (\textbf{Row \textcircled{\scriptsize 15}}).



\vspace{-0.8em}
\begin{algorithm}
\scriptsize
\caption{Largest LE and LER Loss Calculation}
\label{alg:MRLD}
\begin{algorithmic}[1]
\REQUIRE Feature sequence $x \in \mathbb{R}^{T \times C}$, $m=10$, $\tau=1$
\ENSURE Lyapunov exponent $\lambda$ and LER loss $\mathcal{L}_{\text{Lyap}}$
\STATE Compute $M \gets T - (m-1)\tau$, $d \gets m \cdot C$
\FOR{$i=0$ to $M-1$}
    \STATE $E_V[i] \gets [x[i,:], x[i+\tau,:], \dots, x[i+(m-1)\tau,:]]$
\ENDFOR
\FOR{$i,j \in \{0,\dots,M-1\}$}
    \STATE $D[i,j] \leftarrow \lVert E_V[i] - E_V[j] \rVert_2$
\ENDFOR
\FOR{$i=0$ to $M-1$}
    \STATE Find $j^*(i)$ minimizing $D[i,j]$ (exclude Theiler window)
    \FOR{$k=1$ to $k_{\max}$}
        \STATE $d_k[i] \leftarrow \lVert E_V[i+k] - E_V[j^*(i)+k] \rVert_2$
    \ENDFOR
\ENDFOR
\STATE $S(k) \gets \frac{1}{N_k} \sum_i \log(d_k[i]+\epsilon_1)$, Center $k$ and $S(k)$ to get $\tilde{k}, \tilde{S}$
\STATE Calculate $\lambda \gets \frac{\sum \tilde{k}_i \tilde{S}_i}{\sum \tilde{k}_i^2 + \epsilon_2}$ and  $\mathcal{L}_{\text{LER}} \gets |\lambda^{(\hat{Y}_S)} - \lambda^{(Y_V)}|$
\STATE \Return $\lambda$, $\mathcal{L}_{\text{LER}}$
\end{algorithmic}
\end{algorithm}

\vspace{-0.4em}
\subsubsection{\textbf{Multi-scale Detrended Fluctuation Analysis (MSDFA)}}
\vspace{-0.4em}

We introduce Detrended Fluctuation Analysis (DFA) \cite{peng1994mosaic} to quantify fractal-like, long-range temporal correlations that conventional spectrogram losses overlook. Therefore, by computing the root-mean-square fluctuations, the MSDFA loss establishes long-range correlations and fractal scaling behaviors in EMG features.  If omitted, the reconstructed audio leads to muffled prosody even when other losses are present.


For each channel $c \in \{1, \dots,C\}$, we compute the mean $\mu_c$ and define the profile (integrated features) as the cumulative sum of deviations from the mean $y[k,c]$ where $k \in \{1,\dots,T\}$ and $T$ is the number of time frames (\ul{see Alg.}\ref{alg:MSDFA}). This sum constructs the windowed profile tensor $Y_s$ (\textbf{Rows \textcircled{\scriptsize 1}--\textcircled{\scriptsize 3}}). The design matrix $X_s \in \mathbb{R}^{s \times 2}$ is defined with rows $[1, \tau_j]$ where $\tau_j$=$(j-1)/(s-1)$ represents the normalized time within the segment for $j \in \{1, \dots, s\}$ and the projection matrix onto the column space of $X_s$ is $P_s$=$X_s(X_s^T X_s)^{-1} X_s^T \in \mathbb{R}^{s \times s}$. The detrending kernel can be denoted by $K_s$=$I_s - P_s \in \mathbb{R}^{s \times s}$ where $I_s \in \mathbb{R}^{s \times s}$ is the identity matrix. The detrended residuals are computed as $R_s[c, i, j]$=$\sum_{j'=1}^s Y_s[c, i, j'] \cdot K_s[j', j]$ (\textbf{Rows \textcircled{\scriptsize 4}--\textcircled{\scriptsize 7}}). Here, $i \in \{1, \dots, N_s\}$ and $j \in \{1, \dots, s\}$ are segment (window) index and position within the segment, respectively. The MSDFA scaling component for channel $c$ is calculated using the \textbf{Row\textcircled{\scriptsize 8}}, where $\epsilon_3=10^{-8}$ prevents division by zero. $\log s[i]$ and $\log F[c,i]$ are centered logarithmic values, which is RMS across all $R_s$. $s$ and $F$ are pre-defined set of scales for MSDFA and fluctuation function. Given, temporally aligned $\hat{Y}_S[j]$ and $Y_V[i]$, we define the MSDFA loss as $\mathcal{L}_{\text{MSDFA}}(\hat{Y}_S[j], Y_V[i]) = 
 \frac{1}{|\mathcal{V}|} \sum_{c \in \mathcal{V}} \big\vert \alpha_c^{(\hat{Y}_S[j])} - \alpha_c^{(Y_V[i])} \big\vert$, where $|\mathcal{V}|$ is cardinality of the valid channel set.
 


\vspace{-1em}
\begin{algorithm}
\scriptsize
\caption{MSDFA Loss Computation}
\label{alg:MSDFA}
\begin{algorithmic}[1]
\REQUIRE $x \in \mathbb{R}^{T \times C}$, scales $s$
\FOR{$c=1$ to $C$}
    \STATE $\mu_c \leftarrow \text{mean}(x[:,c])$
    \STATE $Y_s \leftarrow \text{cumsum}(x[:,c]-\mu_c)$
    \STATE Construct $X_s$ with rows $[1,\tau_j]$, $\tau_j=(j-1)/(s-1)$
    \STATE $P_s \leftarrow X_s(X_s^T X_s)^{-1}X_s^T$, \quad $K_s \leftarrow I_s-P_s$
    \STATE $R_s \leftarrow Y_s K_s$  \text (detrended residuals)
    \STATE Compute $F[c,i]$ as RMS over $R_s$ for each scale
    \STATE $\alpha_c \leftarrow \frac{\sum \log s[i]\log F[c,i]}{\sum \log s[i]}^2+\epsilon_3$
\ENDFOR
\STATE $\mathcal{L}_{\text{MSDFA}} \leftarrow \frac{1}{|\mathcal{V}|}\sum_{c\in\mathcal{V}} |\alpha_c^{(\hat{Y}_S)}-\alpha_c^{(Y_V)}|$
\end{algorithmic}
\end{algorithm}
\vspace{0em}

\vspace{-0.2em}
\subsection{Audio-level Alignment -- Post-Vocoder Alignment}
\vspace{-0.2em}

\textit{We are the first to apply Post-Vocoder Alignment (PVA) to temporally align generated waveforms with target audio.} Pre-trained vocoders in most baseline methods \cite{gaddy-klein-2020-digital, gaddy-klein-2021-improved} produce misaligned audio due to frame count discrepancies, which prevent frame-level metric computation, such as LSD, STOI, and PESQ. Our approach establishes temporal correspondence through DTW-based alignment in the generated audios. We extract Short-time Fourier Transform (STFT) magnitude spectrograms $S_g$ and $S_t$ from generated and target audio, respectively, normalize each frame, and construct a cosine distance matrix $C$=$\mathbf{1} - \hat{S}_g \hat{S}_t^T$, where $\hat{S}_g$ and $\hat{S}_t$ are normalized values. The optimal alignment path obtained via DTW, is applied to align the vocoder-synthesized audio, which is then reconstructed through inverse STFT for evaluation. Hence, this alignment enables the calculation of the frame-level metrics introduced in this work.

\vspace{-0.8em}
\section{Experiments}
\label{sec:experiment}
\vspace{-0.4em}

\subsection{Dataset and Preprocessing}
\label{dataset_preprocess}
\vspace{-0.4em}

Our training is performed on the state-of-the-art (SOTA) open-vocabulary dataset introduced in \cite{gaddy-klein-2020-digital}. It comprises 19 hours of facial EMG data collected from one English speaker across silent and vocalized speech conditions. Raw EMG signals undergo minimal preprocessing as described in \cite{gaddy-klein-2021-improved}.

\vspace{-01.0em}
\subsection{Training, Hardware, and Hyperparameters}
\vspace{-0.4em}

We train for 80 epochs with batch size 32 using adaptive moment estimation with weight decay (AdamW) optimizer having a learning rate=$10^{-3}$ and weight  decay=$10^{-7}$. Gradients are clipped to norm 0.5. Models are trained in FP32 precision using distributed data parallelism on two NVIDIA RTX 4090 GPUs with AMD Ryzen 7950X3D, 192 GB RAM, \& 10 TB storage.

\vspace{-01.0em}
\subsection{Evaluation Metrics}
\label{subsec:metrics}
\vspace{-0.4em}

For the first time, we introduce comprehensive evaluation of the reconstructed audio using multiple metrics including Log Spectral Distance (LSD) \cite{LSD2022}, Short-time Objective Intelligibility (STOI) \cite{STOI2011}, Perceptual Evaluation of Speech Quality (PESQ) \cite{PESQ2001}, Scale-invariant Signal-to-distortion Ratio (SI-SDR) \cite{SISDR2019}, Non-Intrusive Speech Quality Assessment-Mean Opinion Score (NISQA-MOS) \cite{NISQAMOS2021}, Floating-point Operations (FLOPs), Inference Time (IT) per batch with time steps, $t_s=200$, Word Error Rate (WER), and Real-time Factor (RTF). \textit{There is no work in the literature that uses these extensive metrics for a comprehensive evaluation of the ETS system.}

\vspace{-01.0em} 
\subsection{Total Loss Function}
\vspace{-0.6em}

We employ {\color{black}PD loss ($\mathcal{L}_{PD}$), and phoneme loss ($\mathcal{L}_{Ph}$) from \cite{gaddy-klein-2021-improved}, together with our two novel \textit{chaos-inspired} loss functions, $\mathcal{L}_{LER}$ and $\mathcal{L}_{MSDFA}$. The  $\mathcal{L}_{LER}$ is designed to capture chaotic temporal irregularities, whereas $\mathcal{L}_{MSDFA}$ models chaotic multi-scale dynamics. The encoder generates 80 melspectrogram bins, which are temporally aligned at the frame level using DTW. These aligned bins are then used to compute the LE and to perform MSDFA. The total loss $\mathcal{L}_{\text{total}}$ is: 

\vspace{-2.1em}
{
\begin{equation}
\mathcal{L}_{total} = \mathcal{L}_{PD}+\mathcal{L}_{Ph}+\mathcal{L}_{LER}+\mathcal{L}_{MSDFA}   
\label{eqn:totalloss}
\end{equation}
}
}

\vspace{-02.6em}
\section{Results and Ablation Study}
\vspace{-0.0em}


\vspace{-0.5em}
\subsection{Performance Analysis}
\vspace{-0.4em}


Table \ref{tab:comparison} presents a detailed comparison of CS-ETS against SOTA ETS models. A key distinction of our evaluation is the use of multiple qualitative and quantitative speech-domain metrics that provide rigorous assessment beyond conventional WER. This detailed evaluation is enabled for the first time by our PVA approach, which is absent in the literature, to the best of our knowledge. Previous works \cite{gaddy-klein-2020-digital, gaddy-klein-2021-improved, scheck2023, xie2025neural} report only WER, and none of them do a comprehensive quality assessment using LSD, STOI, PESQ, SI-SDR, and NISQA-MOS due to temporal misalignment issues in vocoder-synthesized audio.

We use three ETS baselines \cite{gaddy-klein-2020-digital,gaddy-klein-2021-improved,scheck2023} in Table \ref{tab:comparison} for comparison. Out of these three baselines, only \cite{gaddy-klein-2020-digital,gaddy-klein-2021-improved} have a reproducible codebase. Therefore, we retrain \cite{gaddy-klein-2020-digital,gaddy-klein-2021-improved} and evaluate all the metrics in Table \ref{tab:comparison} for \cite{gaddy-klein-2020-digital,gaddy-klein-2021-improved}. Table \ref{tab:comparison} indicates that \cite{gaddy-klein-2021-improved} is the best performing model among the baselines in terms of WER, and therefore, we compare our CS-ETS with this best performing model. Table \ref{tab:comparison} indicates that our chaos-inspired CS-ETS delivers the 2.1x best spectral reconstruction (LSD = 1.07 vs 2.25), 4.7x improvement in intelligibility (STOI = 0.61 vs 0.13), 1.25x improvement in noise reduction (SI-SDR = -33.41 vs -41.96) with similar audio quality (i.e., similar PESQ and NISQA-MOS) compared to the best performing baseline  \cite{gaddy-klein-2021-improved}.

\vspace{-0.6em}
\begin{table}[ht]
    \centering
    \caption{Comparative Performance Against SOTA Methods.}
    \vspace{-1em}
    \label{tab:comparison}
    \resizebox{\columnwidth}{!}{%
    \begin{tabular}{l l c c c c c c c c c c} 
        \hline
        & \textbf{Model} & \textbf{LSD$\downarrow$} & \textbf{STOI$\uparrow$} & \textbf{PESQ$\uparrow$} & \textbf{SISDR$\uparrow$}& \textbf{NISQA$\uparrow$}& \textbf{WER$\downarrow$} & \textbf{PVA} & \textbf{Param$\downarrow$}\\ 
        \hline
        $P_0$ & {\color{black}Gaddy et al.} \cite{gaddy-klein-2020-digital} (EMNLP 2020) & 3.23 & 0.04 & 1.17 & -52.52 & 3.28 & 68.00\% & $\times$ & 60.21M\\
        $P_1$ & Gaddy et al. \cite{gaddy-klein-2021-improved} (ACL 2021) & 2.25 & 0.13 & 1.20 & -41.96 & 3.30 & 42.20\% & $\times$ & 54.10M \\
        $P_2$ & Scheck et al. \cite{scheck2023} (ITG 2023) & $\times$ & $\times$ & $\times$ & $\times$ & $\times$ & 46.14\% & $\times$ & $\times$\\
        $P_3$ & \textbf{Proposed CS-ETS (2026)} & \textbf{1.07} & \textbf{0.61} & \textbf{1.19} & \textbf{-33.41} & \textbf{3.31} & \textbf{41.26\%} & $\checkmark$ & \textbf{32.03M}\\ 
        \hline
    \end{tabular}
    }
\end{table}
\vspace{-0.6em}


Please note that CS-ETS achieves superior speech reconstruction compared to \cite{gaddy-klein-2021-improved} using only 59.21\% of the parameters of \cite{gaddy-klein-2021-improved}, achieving substantial model compression while maintaining improved WER and avoiding catastrophic performance degradation. This corresponds to a total model size of 32.03M vs. 54.10M for the baseline \cite{gaddy-klein-2021-improved}, a 40.79\% reduction. \textit{These results indicate that our chaos-inspired Mamba-SWA-MLP encoder captures temporal dependencies with 40.79\% less parameters than the Transformer-based encoder of \cite{gaddy-klein-2021-improved}. Moreover, most recent SOTA works \cite{scheck2023, ren2024, xie2025neural} do not report model size or computational complexity, and their codebases are not publicly available, precluding direct efficiency comparisons.} Therefore, CS-ETS is the first model that is proven to be suitable for 40.79\% fewer parameters, delivering intelligible audio quality across five perceptual metrics with considerably lower parameters and computational complexity (\textit{see Tables \ref{tab:ablation2} and \ref{tab:ablation3} for ablation details}). This proves our important point that explicitly integrating non-linear chaotic physics into models can give better performance with a smaller model size.

\vspace{-0.9em}
\subsection{Computational Efficiency}
\label{Computational Efficiency}
\vspace{-0.5em}


Table \ref{tab:ablation2} compares the computational efficiency of our proposed CS-ETS with the baseline \cite{gaddy-klein-2021-improved}, as \cite{gaddy-klein-2021-improved} is the best performing reproducible model. CS-ETS achieves considerable encoder parameter reduction while maintaining real-time inference. The encoder size drops from 44.03M to 21.96M parameters—a 50.1\% reduction. Our model also shows improved computational efficiency beyond just parameter reduction. CS-ETS lowers FLOPs from 2.40G to 2.08G (13.33\% reduction) and inference time from 5.97ms to 5.55ms per batch, while both models maintain identical RTF (0.0057) for real-time processing. These gains come with better audio quality and intelligibility \textit{(see Table \ref{tab:ablation1})}. \textit{The identical CFE size (10.07M) across both models isolates the encoder as the main source of efficiency improvement, which confirms that LER and MSDFA losses with the Mamba-based SWA offer a more compact yet effective representation for ETS mapping than multi-head self-attention mechanisms.}

\vspace{-0.5em}
\begin{table}[ht!]
    \centering
    \caption{Study on the Computational Efficiency}
    \vspace{-0.9em}
    \label{tab:ablation2}
    \resizebox{\columnwidth}{!}{%
    \begin{tabular}{l l c c c c c c} 
        \hline
        & \textbf{Model} & \textbf{CFE Params} & \textbf{Encoder Params} & \textbf{Total Params} & \textbf{Inference} & \textbf{FLOPs} & \textbf{RTF}\\ \hline
        $P_0$ & Gaddy et al. \cite{gaddy-klein-2021-improved} & 10.07M & 44.03M & 54.10M & 5.97ms & 2.40G & 0.0057 \\ 
        $P_1$ & \textbf{CS-ETS (proposed)} & \textbf{10.07M} & \textbf{21.96M} & \textbf{32.03M} & \textbf{5.55ms} & \textbf{2.08G} & \textbf{0.0057} \\ \hline
    \end{tabular}}
    \vspace{-0.9em}
\end{table}

\vspace{-0.9em}
\subsection{Ablation Study}
\label{ablation_study}
\vspace{-0.5em}

\textbf{a) Chaotic Loss Assessment:} We perform ablation studies on the LER and MSDFA loss components to quantify their individual impact on audio quality in Table \ref{tab:ablation1}. We retrain the baseline model \cite{gaddy-klein-2021-improved} to enable direct comparison with our proposed model across all metrics. All configurations are trained with the same data and optimization settings from Section \ref{sec:experiment}. 

\vspace{-0.7em}
\begin{table}[ht!]
    \centering
    \caption{Ablation Study on Audio Quality Assessment}
    \vspace{-0.9em}
    \label{tab:ablation1}
    \resizebox{\columnwidth}{!}{%
    \begin{tabular}{l l c c c c c c c} 
        \hline
        & \textbf{Model Configuration} & \textbf{LSD} & \textbf{STOI} & \textbf{PESQ} & \textbf{SISDR} & \textbf{NISQA-MOS}& \textbf{WER} & \textbf{Params}\\ \hline
        $P_0$ & Gaddy et al. \cite{gaddy-klein-2021-improved} & 2.25 & 0.13 & 1.20 & -41.96 & 3.30 & 42.20\% & 54.10M\\
        \hline
        $P_1$ & CS-ETS w/o PVA+LER+MSDFA & 1.92 & 0.15 & 1.18 & -36.90 & 3.29 & 50.37\% & 32.03M\\
        $P_2$ & CS-ETS w PVA w/o LER+MSDFA & 1.10 & 0.61 & 1.19 & -31.30 & 3.31 & 50.37\% & 32.03M\\
        $P_3$ & CS-ETS w PVA w/o MSDFA & 1.09 & 0.60 & 1.17 & -32.96 & 3.21 & 49.45\% & 32.03M\\
        $P_4$ & CS-ETS w PVA w/o LER & 1.09 & 0.61 & 1.19 & -32.47 & 3.29 & 48.65\% & 32.03M\\ \hline
        $P_5$ & \textbf{CS-ETS w PVA+LER+MSDFA (proposed)} & \textbf{1.07} & \textbf{0.61} & \textbf{1.19} & \textbf{-33.41} & \textbf{3.31} & \textbf{41.26\%} & 32.03M\\ \hline
    \end{tabular}}
    \vspace{-0.7em}
\end{table}

Row $P_0$ shows the baseline \cite{gaddy-klein-2021-improved} after re-training. \textit{This model yields unreliable audio quality metrics (LSD: 2.25, STOI: 0.13, SISDR: -41.96) without our PVA approach due to frame count mismatches between vocoder-generated and target audio.} This fundamental limitation in all prior work prevented frame-by-frame metrics (LSD, STOI, PESQ in Section \ref{subsec:metrics}) from being computed in their original studies. {\color{black} Row $P_1$ shows that the performance of CS-ETS degrades when we do not have any chaotic losses and PVA in our model. Introducing PVA in row $P_2$ improves frame-level metrics, yielding lower LSD and higher STOI. Also, row $P_2$ shows the impact of the proposed chaos-inspired loss functions, as removing both $\mathcal{L}_{LER}$ and $\mathcal{L}_{MSDFA}$ results in degraded WER.} The critical contribution of our chaotic loss formulation becomes evident in rows $P_4$ and $P_3$: incorporating LER alone reduces WER to 48.65\% while improving spectral quality metrics, and adding MSDFA alone achieves 49.45\% WER. The full model $P_5$ achieves optimal performance across all metrics, which suggests that our newly introduced chaotic losses, LER and MSDFA, provide complementary supervision that synergistically improves both intelligibility and audio quality with a 40.79\% smaller model size. The chaotic losses may provide better temporal stability and multi-scale spectral regularization to overcome fundamental limitations in prior ETS models that relied solely on reconstruction losses.


\noindent \textbf{b) Channel Dimension in CFE Block:} Table \ref{tab:ablation3} shows that CFE channel dimension $C_D=128$ yields poor WER (50.12\%) and quality (NISQA-MOS: 3.08), which indicates insufficient capacity. $C_D=256$ (row $P_1$) reduces WER to 45.53\% and improves metrics. Higher dimensions (512, 768) further enhance quality with NISQA-MOS improving to 3.31, while spectral metrics plateau beyond 256. 

\vspace{-0.7em}
\begin{table}[ht]
    \centering
    \caption{Ablation Study on Channel Dimension in CFE Block}
    \vspace{-0.9em}
    \label{tab:ablation3}
    \resizebox{\columnwidth}{!}{%
    \begin{tabular}{l l c c c c c c} 
        \hline
        & \textbf{Model Configuration} & \textbf{LSD} & \textbf{STOI} & \textbf{PESQ} & \textbf{SISDR} & \textbf{NISQA-MOS}& \textbf{WER}\\ \hline
        $P_0$ & CS-ETS w $C_D=128$ & 1.09 & 0.59 & 1.16 & -32.44 & 3.08 & 50.12\% \\
        $P_1$ & CS-ETS w $C_D=256$ & 1.07 & 0.61 & 1.19 & -31.87 & 3.24 & 45.53\% \\
        $P_2$ & CS-ETS w $C_D=512$ & 1.07 & 0.61 & 1.19 & -33.58 & 3.25 & 44.18\%\\
        $P_3$ & CS-ETS w $C_D=768$ & 1.07 & 0.61 & 1.19 & -33.41 & 3.31 & 41.26\% \\ \hline
    \end{tabular}}
    \vspace{-0.7em}
\end{table}

We use $C_D=768$ (9.6$\times$ mel expansion) for optimal intelligibility and subtle perceptual attributes. The intuition behind the superior performance at 768 dimensions includes: 768 facilitates encoder feature propagation, enables multi-channel EMG disentanglement, and provides sufficient capacity for complex ETS synthesis.

\noindent \textbf{c) Layer Count in Mamba-SWA-MLP:}  We conduct an ablation study on the number of Mamba-SWA-MLP layers in the encoder, varying from 2 to 5 layers. The 4-layer configuration (21.96M encoder parameters) achieves the strongest overall performance with the lowest LSD (1.07), highest STOI (0.61), PESQ (1.19), and NISQA-MOS (3.31), alongside the lowest WER (41.26\%). In contrast, the 2-layer model underperforms across most metrics, while the 5-layer model exhibits degraded performance across LSD, PESQ, NISQA-MOS, and WER despite a substantial increase in encoder parameters.

\vspace{-0.5em}
\begin{table}[ht]
    \centering
    \caption{Ablation Study on Layer Count in Mamba-SWA-MLP}
    \vspace{-0.8em}
    \label{tab:ablation4}
    \resizebox{\columnwidth}{!}{%
    \begin{tabular}{l l c c c c c c c} 
        \hline
        & \textbf{Encoder} & \textbf{Encoder Params} & \textbf{LSD} & \textbf{STOI} & \textbf{PESQ} & \textbf{SISDR} & \textbf{NISQA-MOS}& \textbf{WER}\\ \hline
        $P_0$ & Layers=2 & 12.75M & 1.11 & 0.59 & 1.17 & -33.15 & 3.23 & 56.86\% \\
        $P_1$ & Layers=3 & 19.13M & 1.09 & 0.60 & 1.19 & -31.51 & 3.28 & 50.12\% \\
        $P_2$ & Layers=4 & 21.96M & 1.07 & 0.61 & 1.19 & -33.41 & 3.31 & 41.26\% \\
        $P_3$ & Layers=5 & 31.88M & 1.08 & 0.61 & 1.18 & -33.15 & 3.26 & 43.93\% \\ \hline
    \end{tabular}}
    \vspace{-0.7em}
\end{table}

\vspace{-0.9em}
\section{Subjective Analysis}
\label{subsec:Subjective Test}
\vspace{-0.4em}

Subjective comparison of CS-ETS against \cite{gaddy-klein-2021-improved} and unprocessed EMG is conducted by a panel of 10 persons. We use 5-point (1=bad to 5=excellent) Mean Opinion Score (MOS) ratings for subjective evaluation. The unprocessed EMG has MOS=1, meaning completely unrecognizable. \textit{Our CS-ETS, having MOS=4.21, outperforms \cite{gaddy-klein-2021-improved}, which has MOS=3.98.  From this result, we can comment that CS-ETS may better reconstruct missing chaotic speech artifacts from the raw EMG data that is neglected by the previous work.} Results provide strong evidence that CS-ETS consistently generates perceptually higher quality audio, favored by a wide range of listeners.

\vspace{-0.8em}
\section{Conclusion and Limitations}
\vspace{-0.4em}

This paper introduces CS-ETS which is the first chaos-inspired Samba-based ETS architecture. Our ``modified Samba'' encoder with novel chaos-inspired losses achieves better performance at substantially lower computational cost. We propose post-vocoder alignment that allows comprehensive perceptual evaluation (LSD, STOI, SISDR, NISQA-MOS) that is previously absent in ETS research. \textit{Experiments on Gaddy and Klein datasets \cite{gaddy-klein-2020-digital} validate that chaos theory improves articulatory-to-acoustic mapping with 40.79\% fewer parameters and 13.33\% reduction in FLOPS, establishing for the first time that chaos-inspired Samba-based architectures are promising for ETS generation.} However, we do not use multiple datasets in a noisy setting in this paper. We will handle these in our upcoming work.

\vspace{-0.900em}
\section{Generative AI Use Disclosure}
\vspace{-0.3em}

We acknowledge the use of \textit{Elicit} and \textit{ChatGPT} during early-stage brainstorming to explore theories and help structure the architectural design, as well as for proofreading and language polishing of manuscript drafts. All scientific concepts, methodological decisions, experimental design, implementations, results, and final wording were developed by the authors and independently verified by the authors.

\bibliographystyle{IEEEtran}
\bibliography{mybib}

@inproceedings{gaddy-klein-2020-digital,
    title = "Digital Voicing of Silent Speech",
    author = "Gaddy, David  and
      Klein, Dan",
    booktitle = "Proceedings of the 2020 Conference on Empirical Methods in Natural Language Processing (EMNLP)",
    month = nov,
    year = "2020",
    address = "Online",
    publisher = "Association for Computational Linguistics",
    url = "https://aclanthology.org/2020.emnlp-main.445/",
    doi = "10.18653/v1/2020.emnlp-main.445",
    pages = "5521--5530"
}

@article{Theiler1986,
  author  = {James Theiler},
  title   = {Spurious Dimension from Correlation Algorithms Applied to Limited Time-Series Data},
  journal = {Physical Review A},
  volume  = {34},
  number  = {3},
  pages   = {2427--2432},
  year    = {1986},
  doi     = {10.1103/PhysRevA.34.2427}
}

@article{oseledec1968multiplicative,
  title={A multiplicative ergodic theorem: Lyapunov characteristic numbers for dynamical systems},
  author={Oseledec, V.I.},
  journal={Trans. Moscow Math. Soc.},
  volume={19},
  pages={197--231},
  year={1968}
}

@article{wolf1985determining,
  title={Determining Lyapunov exponents from a time series},
  author={Wolf, Alan and Swift, J. B. and Swinney, Harry L. and Vastano, John A.},
  journal={Physica D: Nonlinear Phenomena},
  volume={16},
  number={3},
  pages={285--317},
  year={1985},
  publisher={Elsevier}
}

@article{peng1994mosaic,
  title={Mosaic organization of DNA nucleotides},
  author={Peng, C.-K. and Buldyrev, S. V. and Havlin, S. and Simons, M. and Stanley, H. E. and Goldberger, A. L.},
  journal={Physical Review E},
  volume={49},
  number={2},
  pages={1685--1689},
  year={1994},
  publisher={American Physical Society},
  doi={10.1103/PhysRevE.49.1685}
}

@misc{tamiti2025nonlinearframeworkspeechbandwidth,
      title={Nonlinear Framework for Speech Bandwidth Extension}, 
      author={Tarikul Islam Tamiti and Nursad Mamun and Anomadarshi Barua},
      year={2025},
      eprint={2507.15970},
      archivePrefix={arXiv},
      primaryClass={cs.SD},
      url={https://arxiv.org/abs/2507.15970}, 
}

@misc{tamiti2025nldsibwenonlineardynamical,
      title={NLDSI-BWE: Non Linear Dynamical Systems-Inspired Multi Resolution Discriminators for Speech Bandwidth Extension}, 
      author={Tarikul Islam Tamiti and Anomadarshi Barua},
      year={2025},
      eprint={2510.01109},
      archivePrefix={arXiv},
      primaryClass={cs.SD},
      url={https://arxiv.org/abs/2510.01109}, 
}

@article{little2007exploiting,
  title={Exploiting nonlinear recurrence and fractal scaling properties for voice disorder detection},
  author={Little, Max and Mcsharry, Patrick and Roberts, Stephen and Costello, Declan and Moroz, Irene},
  journal={Nature Precedings},
  pages={1--1},
  year={2007},
  publisher={Nature Publishing Group UK London}
}

@article{samba2024,
  title={Samba: Simple hybrid state space models for efficient unlimited context language modeling},
  author={Ren, Liliang and Liu, Yang and Lu, Yadong and Shen, Yelong and Liang, Chen and Chen, Weizhu},
  journal={arXiv preprint arXiv:2406.07522},
  year={2024}
}

@misc{chaos2025,
Author = {Tarikul Islam Tamiti and Nursad Mamun and Anomadarshi Barua},
Title = {Nonlinear Framework for Speech Bandwidth Extension},
Year = {2025},
Eprint = {arXiv:2507.15970},
}

@article{hifigan2020,
  title={Hifi-gan: Generative adversarial networks for efficient and high fidelity speech synthesis},
  author={Kong, Jungil and Kim, Jaehyeon and Bae, Jaekyoung},
  journal={Advances in neural information processing systems},
  volume={33},
  pages={17022--17033},
  year={2020}
}

@inproceedings{mamba2024,
  title={Mamba: Linear-time sequence modeling with selective state spaces},
  author={Gu, Albert and Dao, Tri},
  booktitle={First conference on language modeling},
  year={2024}
}

@article{swa2020,
  title={Longformer: The long-document transformer},
  author={Beltagy, Iz and Peters, Matthew E and Cohan, Arman},
  journal={arXiv preprint arXiv:2004.05150},
  year={2020}
}

@article{LSD2022,
  title={Neural vocoder is all you need for speech super-resolution},
  author={Liu, Haohe and Choi, Woosung and Liu, Xubo and Kong, Qiuqiang and Tian, Qiao and Wang, DeLiang},
  journal={arXiv preprint arXiv:2203.14941},
  year={2022}
}

@article{STOI2011,
  title={An algorithm for intelligibility prediction of time--frequency weighted noisy speech},
  author={Taal, Cees H and Hendriks, Richard C and Heusdens, Richard and Jensen, Jesper},
  journal={IEEE Transactions on audio, speech, and language processing},
  volume={19},
  number={7},
  pages={2125--2136},
  year={2011},
  publisher={IEEE}
}

@inproceedings{PESQ2001,
  title={Perceptual evaluation of speech quality (PESQ)-a new method for speech quality assessment of telephone networks and codecs},
  author={Rix, Antony W and Beerends, John G and Hollier, Michael P and Hekstra, Andries P},
  booktitle={2001 IEEE international conference on acoustics, speech, and signal processing. Proceedings (Cat. No. 01CH37221)},
  volume={2},
  pages={749--752},
  year={2001},
  organization={IEEE}
}

@inproceedings{SISDR2019,
  title={SDR--half-baked or well done?},
  author={Le Roux, Jonathan and Wisdom, Scott and Erdogan, Hakan and Hershey, John R},
  booktitle={ICASSP 2019-2019 IEEE International Conference on Acoustics, Speech and Signal Processing (ICASSP)},
  pages={626--630},
  year={2019},
  organization={IEEE}
}

@article{NISQAMOS2021,
  title={NISQA: A deep CNN-self-attention model for multidimensional speech quality prediction with crowdsourced datasets},
  author={Mittag, Gabriel and Naderi, Babak and Chehadi, Assmaa and M{\"o}ller, Sebastian},
  journal={arXiv preprint arXiv:2104.09494},
  year={2021}
}

@inproceedings{scheck2023,
  title={Stream-ETS: Low-latency end-to-end speech synthesis from electromyography signals},
  author={Scheck, Kevin and Ivucic, Darius and Ren, Zhao and Schultz, Tanja},
  booktitle={Speech Communication; 15th ITG Conference},
  pages={200--204},
  year={2023},
  organization={VDE}
}

@inproceedings{ren2024,
  title={Diff-ets: Learning a diffusion probabilistic model for electromyography-to-speech conversion},
  author={Ren, Zhao and Scheck, Kevin and Hou, Qinhan and van Gogh, Stefano and Wand, Michael and Schultz, Tanja},
  booktitle={2024 46th Annual International Conference of the IEEE Engineering in Medicine and Biology Society (EMBC)},
  pages={1--4},
  year={2024},
  organization={IEEE}
}

@article{satterlee2025,
  title={Sentence-Level Silent Speech Recognition Using a Wearable EMG/EEG Sensor System with AI-Driven Sensor Fusion and Language Model},
  author={Satterlee, Nicholas and Zuo, Xiaowei and Moon, Kee and Lee, Sung Q and Peterson, Matthew and Kang, John S},
  journal={Sensors},
  volume={25},
  number={19},
  pages={6168},
  year={2025},
  publisher={MDPI}
}

@article{dong2023,
  title={Decoding silent speech commands from articulatory movements through soft magnetic skin and machine learning},
  author={Dong, Penghao and Li, Yizong and Chen, Si and Grafstein, Justin T and Khan, Irfaan and Yao, Shanshan},
  journal={Materials Horizons},
  volume={10},
  number={12},
  pages={5607--5620},
  year={2023},
  publisher={Royal Society of Chemistry}
}

@article{kurotaki2025,
  title={Soft Active EMG Interface for Machine Learning-Enabled Silent Speech Recognition},
  author={Kurotaki, Yuta and Yamakoshi, Shunsuke and Yoshida, Reitaro and Isoda, Yutaka and Takano, Tamami and Isano, Yuji and Miyake, Yusuke and Kuribayashi, Kentaro and Ota, Hiroki},
  year={2025}
}

@article{enoka2001,
  title={Motor unit physiology: some unresolved issues},
  author={Enoka, Roger M and Fuglevand, Andrew J},
  journal={Muscle \& Nerve: Official Journal of the American Association of Electrodiagnostic Medicine},
  volume={24},
  number={1},
  pages={4--17},
  year={2001},
  publisher={Wiley Online Library}
}

@article{farina2015,
  title={Common synaptic input to motor neurons, motor unit synchronization, and force control},
  author={Farina, Dario and Negro, Francesco},
  journal={Exercise and sport sciences reviews},
  volume={43},
  number={1},
  pages={23--33},
  year={2015},
  publisher={LWW}
}

@article{keenan2005,
  title={Influence of amplitude cancellation on the simulated surface electromyogram},
  author={Keenan, Kevin G and Farina, Dario and Maluf, Katrina S and Merletti, Roberto and Enoka, Roger M},
  journal={Journal of applied physiology},
  volume={98},
  number={1},
  pages={120--131},
  year={2005},
  publisher={American Physiological Society}
}

@article{saltzman1989,
  title={A dynamical approach to gestural patterning in speech production},
  author={Saltzman, Elliot L and Munhall, Kevin G},
  journal={Ecological psychology},
  volume={1},
  number={4},
  pages={333--382},
  year={1989},
  publisher={Taylor \& Francis}
}

@article{tuller1984,
  title={for Coordinative Structures},
  author={Tuller, JA Scott Kelso Betty and Fowler, E Vatikiotis-Bateson Carol A},
  journal={journal of Experimental Psychology},
  volume={10},
  number={6},
  pages={812--832},
  year={1984}
}

@inproceedings{gaddy-klein-2021-improved,
    title = "An Improved Model for Voicing Silent Speech",
    author = "Gaddy, David and Klein, Dan",
    booktitle = "Proceedings of the 59th Annual Meeting of the Association for Computational Linguistics and the 11th International Joint Conference on Natural Language Processing (Volume 2: Short Papers)",
    month = aug,
    year = "2021",
    address = "Online",
    publisher = "Association for Computational Linguistics",
    url = "https://aclanthology.org/2021.acl-short.23/",
    doi = "10.18653/v1/2021.acl-short.23",
    pages = "175--181"
}

@article{xie2025neural,
  title={Neural Chinese silent speech recognition with facial electromyography},
  author={Xie, Liang and Zhang, Yakun and Yuan, Hao and Zhang, Meishan and Zhang, Xingyu and Zheng, Changyan and Yan, Ye and Yin, Erwei},
  journal={Speech Communication},
  volume={171},
  pages={103230},
  year={2025},
  publisher={Elsevier}
}

\end{document}